\documentclass[12pt,aps,prd,showpacs,amsmath,amssymb]{revtex4}
\usepackage{graphicx}
\usepackage{epstopdf}
\usepackage{multirow}
\usepackage{subfigure}
\usepackage{multirow}
\usepackage{appendix}
\input epsf
\begin{document}
\title{P-wave $\Omega_{b}$ states: masses and pole residues}
\author{Yong-Jiang Xu\footnote{xuyongjiang13@nudt.edu.cn}, Yong-Lu Liu, and Ming-Qiu Huang\footnote{corresponding author: mqhuang@nudt.edu.cn}}
\affiliation{Department of Physics, College of Liberal Arts and Sciences, National University of Defense Technology , Changsha, 410073, Hunan, China}
\date{}
\begin{abstract}
In this paper, we consider all P-wave $\Omega_{b}$ states represented by interpolating currents with a derivative and calculate the corresponding masses and pole residues with the method of QCD sum rule. Due to the large uncertainties in our calculation compared with the small difference in the masses of the excited $\Omega_{b}$ states observed by the LHCb collaboration, it is necessary to study other properties of the P-wave $\Omega_{b}$ states represented by the interpolating currents investigated in the present work in order to have a better understanding about the four excited $\Omega_{b}$ states observed by the LHCb collaboration.
\end{abstract}
\maketitle

\section{Introduction}\label{sec1}

In 2017, the LHCb collaboration observed five narrow excited $\Omega_{c}$ states, $\Omega_{c}(3000)$, $\Omega_{c}(3050)$, $\Omega_{c}(3066)$, $\Omega_{c}(3090)$ and $\Omega_{c}(3119)$, in the $\Xi^{+}_{c}K^{-}$ mass spectrum \cite{lhcb1}. Recently, they reported four excited $\Omega_{b}$ states in the $\Xi^{0}_{b}K^{-}$ mass spectrum \cite{lhcb2},
\begin{eqnarray}
&&\Omega_{b}(6316): m=6315.64\pm0.31\pm0.07\pm0.50\mbox{MeV},\Gamma<2.8\mbox{MeV},\nonumber\\
&&\Omega_{b}(6330): m=6330.30\pm0.28\pm0.07\pm0.50\mbox{MeV},\Gamma<3.1\mbox{MeV},\nonumber\\
&&\Omega_{b}(6340): m=6339.71\pm0.26\pm0.05\pm0.50\mbox{MeV},\Gamma<1.5\mbox{MeV},\nonumber\\
&&\Omega_{b}(6350): m=6349.88\pm0.35\pm0.05\pm0.50\mbox{MeV},\Gamma=1.4^{+1.0}_{-0.8}\pm0.1\mbox{MeV}.
\end{eqnarray}
Following the experimental progresses, there have been plenty of theoretical works concerning various properties of these excited $\Omega_{Q}$ ($Q=b,c$) states \cite{omegac1,omegac2,omegac3,omegac4,omegac5,omegac6,omegac7,omegac8,omegac9,omegac10,omegac11,omegac12,omegac13,
omegac14,omegac15,omegac16,omegab1,omegab2,omegab3,omegab4,omegab5,omegab6,omegab7,omegab8} and other excited heavy baryons\cite{others1,others2,others3,others4,others5,others6,others7,others8,others9,others10}.

Two kinds of excitations, the $\rho$-mode and $\lambda$-mode, exist in the excited $\Omega_{Q}$ states. The $\rho$-mode excitation is the excitation between two strange quarks, while the $\lambda$-mode one is the excitation between the strange diquark and bottom (charm) quark. In Ref.\cite{chx}, the authors systematically considered all possible baryon currents with a derivative for internal $\rho$- and $\lambda$-mode excitations, and studied the P-wave charmed baryons using the QCD sum rule method in the framework of heavy quark effective theory. In Ref.\cite{omegac10,omegab4}, the authors studied these excited states using the method of QCD sum rule in the framework of QCD.

In this paper, we construct the full QCD counterparts of the interpolating currents considered in Ref.\cite{chx} and study P-wave $\Omega_{b}$ excited states by QCD sum rule method \cite{SVZ}. The basic idea of the QCD sum rule method is that the correlation function of interpolating currents of hadrons can be represented in terms of hadronic parameters (the so-called hadronic side) and calculated at quark-gluon level by operator product expansion (OPE) (the so-called QCD side), and then by matching the two expressions we can extract the physical quantities of the considered hadron.

The rest of the paper is organized as follows. In Sec.\ref{sec2}, we construct the interpolating currents and derive the needed sum rules. Sec.\ref{sec3} is devoted to the numerical analysis and a short summary is given in Sec.\ref{sec4}. In Appendix \ref{appendix2}, OPE results are shown.

\section{The derivation of the sum rules}\label{sec2}

\subsection{The interpolating currents}

According to Ref.\cite{chx}, we introduce the symbols [$\Omega_{b},j_{l},s_{l},\rho/\lambda$] and $J^{\alpha_{1}\alpha_{2}\cdots\alpha_{j-\frac{1}{2}}}_{j,P,\Omega_{b},j_{l},s_{l},\rho/\lambda}$ to denote the P-wave $\Omega_{b}$ multiplets and the interpolating currents, respectively, where $j$ is the total angular momentum, $P$ is the parity, $j_{l}$ and $s_{l}$ are the total angular momentum and spin angular momentum of the light components and $\rho$($\lambda$) denotes the $\rho$($\lambda$)-mode excitations. The general interpolating currents of $\Omega_{b}$ baryons can be written as
\begin{equation}
J(x)\sim\epsilon^{abc}[s^{aT}(x)C\Gamma_{1}s^{b}(x)]\Gamma_{2}b^{c}(c),
\end{equation}
where $a$, $b$, and $c$ are color indices, $\epsilon^{abc}$ is the totally antisymmetric tensor, $C$ is the charge conjugation operator, $T$ denotes the matrix transpose on the Dirac spinor indices, $s(x)$ and $b(x)$ are the strange and bottom quark fields, respectively. The state function corresponding to the diquark $\epsilon^{abc}[s^{aT}(x)C\Gamma_{1}s^{b}(x)]$ can be written as $|color\rangle\otimes|flavor,~spin,~space\rangle$ and should be antisymmetric under the interchange of the two strange quarks. Now, the color part and flavor part are antisymmetric and symmetric, respectively. The spin part is antisymmetric for the scalar diquark $\epsilon^{abc}[s^{aT}(x)C\gamma_{5}s^{b}(x)]$ and symmetric for the axial-vector diquark $\epsilon^{abc}[s^{aT}(x)C\gamma_{\mu}s^{b}(x)]$, respectively. The spatial wave function is antisymmetric and symmetric corresponding to the the $\rho$-mode and $\lambda$-mode excitation, respectively. For example, if the spin angular momentum of the diquark is 0, the excitation in the $\Omega_{b}$ state should be the $\rho$-mode, and we have the baryon-multiplet [$\Omega_{b}$, 1, 0, $\rho$].
Consequently, the P-wave $\Omega_{b}$ states can be classified into four multiplets, [$\Omega_{b}$, 1, 0, $\rho$], [$\Omega_{b}$, 0, 1, $\lambda$], [$\Omega_{b}$, 1, 1, $\lambda$] and [$\Omega_{b}$, 2, 1, $\lambda$], and the corresponding interpolating currents are
\begin{itemize}
  \item{[$\Omega_{b}$, 1, 0, $\rho$]}:
  \begin{eqnarray}\label{interpolating current}
  &&J_{1/2,-,\Omega_{b},1,0,\rho}(x)=i\epsilon_{abc}\{[D_{\mu}(x)s^{T}(x)]^{a}C\gamma_{5}s^{b}(x)-
  s^{T}(x)C\gamma_{5}[D_{\mu}s(x)]^{b}\}\gamma^{\mu}\gamma_{5}b^{c}(x),\nonumber\\
  &&J^{\alpha}_{3/2,-,\Omega_{b},1,0,\rho}(x)=i\epsilon_{abc}\{[D_{\mu}(x)s^{T}(x)]^{a}C\gamma_{5}s^{b}(x)-
  s^{T}(x)C\gamma_{5}[D_{\mu}s(x)]^{b}\}\Gamma^{\alpha\mu}b^{c}(x),
  \end{eqnarray}
  with $\Gamma^{\alpha\mu}=g^{\alpha\mu}-\frac{1}{4}\gamma^{\alpha}\gamma^{\mu}$,
  \item{[$\Omega_{b}$, 0, 1, $\lambda$]}:
  \begin{equation}
  J_{1/2,-,\Omega_{b},0,1,\lambda}(x)=i\epsilon_{abc}\{[D_{\mu}(x)s^{T}(x)]^{a}C\gamma^{\mu}s^{b}(x)+
  s^{T}(x)C\gamma^{\mu}[D_{\mu}s(x)]^{b}\}b^{c}(x),
  \end{equation}
  \item{[$\Omega_{b}$, 1, 1, $\lambda$]}:
  \begin{eqnarray}
  &&J_{1/2,-,\Omega_{b},1,1,\lambda}(x)=i\epsilon_{abc}\{[D_{\mu}(x)s^{T}(x)]^{a}C\gamma_{\nu}s^{b}(x)+
  s^{T}(x)C\gamma_{\nu}[D_{\mu}s(x)]^{b}\}\sigma^{\mu\nu}b^{c}(x),\nonumber\\
  &&J^{\alpha}_{3/2,-,\Omega_{b},1,1,\lambda}(x)=i\epsilon_{abc}\{[D_{\mu}(x)s^{T}(x)]^{a}C\gamma_{\nu}s^{b}(x)+
  s^{T}(x)C\gamma_{\nu}[D_{\mu}s(x)]^{b}\}\Gamma^{\alpha\mu\nu}_{1}b^{c}(x),
  \end{eqnarray}
  with $\Gamma^{\alpha\mu\nu}_{1}=(g^{\alpha\mu}\gamma^{\nu}-g^{\alpha\nu}\gamma^{\mu}-
  \frac{1}{4}\gamma^{\alpha}\gamma^{\mu}\gamma_{\nu}+\frac{1}{4}\gamma^{\alpha}\gamma^{\nu}\gamma_{\mu})\gamma_{5}$,
  \item{[$\Omega_{b}$, 2, 1, $\lambda$]}:
  \begin{eqnarray}
  &&J^{\alpha}_{3/2,-,\Omega_{b},2,1,\lambda}(x)=i\epsilon_{abc}\{[D_{\mu}(x)s^{T}(x)]^{a}C\gamma_{\nu}s^{b}(x)+
  s^{T}(x)C\gamma_{\nu}[D_{\mu}s(x)]^{b}\}\Gamma^{\alpha\mu\nu}_{2}b^{c}(x),\nonumber\\
  &&J^{\alpha_{1}\alpha_{2}}_{5/2,-,\Omega_{b},2,1,\lambda}(x)=i\epsilon_{abc}\{[D_{\mu}(x)s^{T}(x)]^{a}C\gamma_{\nu}s^{b}(x)+
  s^{T}(x)C\gamma_{\nu}[D_{\mu}s(x)]^{b}\}\Gamma^{\alpha_{1}\alpha_{2}\mu\nu}b^{c}(x),
  \end{eqnarray}
  where
  \begin{equation} \Gamma^{\alpha\mu\nu}_{2}=(g^{\alpha\mu}\gamma^{\nu}+g^{\alpha\nu}\gamma^{\mu}-\frac{1}{2}g^{\mu\nu}\gamma^{\alpha})\gamma_{5}, \end{equation}
  \begin{eqnarray}
  \Gamma^{\alpha_{1}\alpha_{2}\mu\nu}=&&g^{\alpha_{1}\mu}g^{\alpha_{2}\nu}+g^{\alpha_{1}\nu}g^{\alpha_{2}\mu}
  -\frac{1}{3}g^{\alpha_{1}\alpha_{2}}g^{\mu\nu}-\frac{1}{6}g^{\alpha_{1}\mu}\gamma^{\alpha_{2}}\gamma^{\nu}\nonumber\\
  &&-\frac{1}{6}g^{\alpha_{1}\nu}\gamma^{\alpha_{2}}\gamma^{\mu}-\frac{1}{6}g^{\alpha_{2}\nu}\gamma^{\alpha_{1}}\gamma^{\mu}
  -\frac{1}{6}g^{\alpha_{2}\mu}\gamma^{\alpha_{1}}\gamma^{nu}.
  \end{eqnarray}
\end{itemize}
In the above equations, $D_{\mu}(x)=\partial_{\mu}-ig_{s}A_{\mu}(x)$ is the gauge-covariant derivative, $a$, $b$, and $c$ are color indices, $C$ is the charge conjugation operator, $T$ denotes the matrix transpose on the Dirac spinor indices, $s(x)$ and $b(x)$ are the strange and bottom quark fields, respectively.

\subsection{The sum rules}

In order to obtain the mass sum rules of the P-wave excited $\Omega_{b}$ states, we begin with the following two-point correlation function of the interpolating currents constructed in the previous subsection,
\begin{equation}\label{2-point correlator}
\Pi^{\alpha_{1}\alpha_{2}\cdots\alpha_{j-\frac{1}{2}}\beta_{1}\beta_{2}\cdots\beta_{j-\frac{1}{2}}}(p)=i\int dx^{4}e^{ipx}\langle0\mid\textsl{T}[J^{\alpha_{1}\alpha_{2}\cdots\alpha_{j-\frac{1}{2}}}_{j,P,\Omega_{b},j_{l},s_{l},\rho/\lambda}(x)
\bar{J}^{\beta_{1}\beta_{2}\cdots\beta_{j-\frac{1}{2}}}_{j,P,\Omega_{b},j_{l},s_{l},\rho/\lambda}(0)]\mid0\rangle.
\end{equation}

Firstly, we should represent phenomenologically the two-point correlation function (\ref{2-point correlator}) in terms of hadronic parameters. To this end, we insert a complete set of states with the same quantum numbers as the interpolating field, perform the integral over space-time coordinates and finally obtain
\begin{eqnarray}
\Pi^{(Phy)\alpha_{1}\cdots\alpha_{j-\frac{1}{2}}\beta_{1}\cdots\beta_{j-\frac{1}{2}}}(p)=&&
\frac{1}{m^{2}_{j,P,\Omega_{b},j_{l},s_{l},\rho/\lambda}-p^{2}}\langle0|J^{\alpha_{1}\cdots\alpha_{j-\frac{1}{2}}}_{j,P,\Omega_{b},j_{l},s_{l},\rho/\lambda}
|j,P,\Omega_{b},j_{l},s_{l},\rho/\lambda,p\rangle\nonumber\\&&\langle j,P,\Omega_{b},j_{l},s_{l},\rho/\lambda,p|
\bar{J}^{\beta_{1}\cdots\beta_{j-\frac{1}{2}}}_{j,P,\Omega_{b},j_{l},s_{l},\rho/\lambda}|0\rangle
+\mbox{higher resonances}.
\end{eqnarray}
We parameterize the matrix element $\langle0|J^{\alpha_{1}\alpha_{2}\cdots\alpha_{j-\frac{1}{2}}}_{j,P,\Omega_{b},j_{l},s_{l},\rho/\lambda}
|j,P,\Omega_{b},j_{l},s_{l},\rho/\lambda,p\rangle$ in terms of the current-hadron coupling constant (pole residue) $f_{j,P,\Omega_{b},j_{l},s_{l},\rho/\lambda}$ and spinor $u^{\alpha_{1}\alpha_{2}\cdots\alpha_{j-\frac{1}{2}}}(p)$,
\begin{equation}
\langle0|J^{\alpha_{1}\alpha_{2}\cdots\alpha_{j-\frac{1}{2}}}_{j,P,\Omega_{b},j_{l},s_{l},\rho/\lambda}
|j,P,\Omega_{b},j_{l},s_{l},\rho/\lambda,p\rangle=f_{j,P,\Omega_{b},j_{l},s_{l},\rho/\lambda}u^{\alpha_{1}\alpha_{2}\cdots\alpha_{j-\frac{1}{2}}}(p).
\end{equation}
As a result, we have,
\begin{itemize}
  \item{for spin-$\frac{1}{2}$ baryon}:
  \begin{equation}\label{hadronic side}
  \Pi^{(Phy)}(p)=\frac{f^{2}_{1/2}}{m^{2}_{1/2}-p^{2}}(\not\!{p}+m_{1/2})+\mbox{higher resonances},
  \end{equation}
  \item{for spin-$\frac{3}{2}$ baryon}:
  \begin{eqnarray}
  \Pi^{(Phy)\alpha_{1}\beta_{1}}(p)=&&\frac{f^{2}_{3/2}}{m^{2}_{3/2}-p^{2}}(\not\!{p}+m_{3/2})(-g^{\alpha_{1}\beta_{1}}
  +\frac{\gamma^{\alpha_{1}}\gamma^{\beta_{1}}}{3}+\frac{2p^{\alpha_{1}}p^{\beta_{1}}}{3m^{2}_{3/2}}
  -\frac{p^{\alpha_{1}}\gamma^{\beta_{1}}-p^{\beta_{1}}\gamma^{\alpha_{1}}}{3m_{3/2}})
  \nonumber\\&&+\mbox{higher resonances},
  \end{eqnarray}
  \item{for spin-$\frac{5}{2}$ baryon}:
  \begin{eqnarray}
  \Pi^{(Phy)\alpha_{1}\alpha_{2}\beta_{1}\beta_{2}}(p)=&&\frac{f^{2}_{5/2}}{m^{2}_{5/2}-p^{2}}(\not\!{p}+m_{5/2})
  [\frac{\tilde{g}^{\alpha_{1}\beta_{1}}\tilde{g}^{\alpha_{2}\beta_{2}}+\tilde{g}^{\alpha_{1}\beta_{2}}\tilde{g}^{\alpha_{2}\beta_{1}}}{2}
  -\frac{\tilde{g}^{\alpha_{1}\alpha_{2}}\tilde{g}^{\beta_{1}\beta_{2}}}{5}\nonumber\\&&-\frac{1}{10}(\gamma^{\alpha_{1}}\gamma^{\beta_{1}}
  +\frac{\gamma^{\alpha_{1}}p^{\beta_{1}}-\gamma^{\beta_{1}}p^{\alpha_{1}}}{m_{5/2}}-\frac{p^{\alpha_{1}}p^{\beta_{1}}}{m^{2}_{5/2}})
  \tilde{g}^{\alpha_{2}\beta_{2}}\nonumber\\&&-\frac{1}{10}(\gamma^{\alpha_{2}}\gamma^{\beta_{1}}
  +\frac{\gamma^{\alpha_{2}}p^{\beta_{1}}-\gamma^{\beta_{1}}p^{\alpha_{2}}}{m_{5/2}}-\frac{p^{\alpha_{2}}p^{\beta_{1}}}{m^{2}_{5/2}})
  \tilde{g}^{\alpha_{1}\beta_{2}}\nonumber\\&&-\frac{1}{10}(\gamma^{\alpha_{1}}\gamma^{\beta_{2}}
  +\frac{\gamma^{\alpha_{1}}p^{\beta_{2}}-\gamma^{\beta_{2}}p^{\alpha_{1}}}{m_{5/2}}-\frac{p^{\alpha_{1}}p^{\beta_{2}}}{m^{2}_{5/2}})
  \tilde{g}^{\alpha_{2}\beta_{1}}\nonumber\\&&-\frac{1}{10}(\gamma^{\alpha_{2}}\gamma^{\beta_{2}}
  +\frac{\gamma^{\alpha_{2}}p^{\beta_{2}}-\gamma^{\beta_{2}}p^{\alpha_{2}}}{m_{5/2}}-\frac{p^{\alpha_{2}}p^{\beta_{2}}}{m^{2}_{5/2}})
  \tilde{g}^{\alpha_{1}\beta_{1}}]\nonumber\\&&+\mbox{higher resonances},
  \end{eqnarray}
\end{itemize}
where we have used the following formulas
\begin{equation}
\sum_{s}u(p,s)\bar{u}(p,s)=\not\!{p}+m_{1/2},
\end{equation}
\begin{equation}
\sum_{s}u^{\alpha_{1}}(p,s)\bar{u}^{\beta_{1}}(p,s)=(\not\!{p}+m_{3/2})(-g^{\alpha_{1}\beta_{1}}
  +\frac{\gamma^{\alpha_{1}}\gamma^{\beta_{1}}}{3}+\frac{2p^{\alpha_{1}}p^{\beta_{1}}}{3m^{2}_{3/2}}
  -\frac{p^{\alpha_{1}}\gamma^{\beta_{1}}-p^{\beta_{1}}\gamma^{\alpha_{1}}}{3m_{3/2}}),
\end{equation}
\begin{eqnarray}
\sum_{s}u^{\alpha_{1}\alpha_{2}}(p,s)\bar{u}^{\beta_{1}\beta_{2}}(p,s)=&&(\not\!{p}+m_{5/2})[\frac{\tilde{g}^{\alpha_{1}\beta_{1}}\tilde{g}^{\alpha_{2}\beta_{2}}+\tilde{g}^{\alpha_{1}\beta_{2}}\tilde{g}^{\alpha_{2}\beta_{1}}}{2}
  -\frac{\tilde{g}^{\alpha_{1}\alpha_{2}}\tilde{g}^{\beta_{1}\beta_{2}}}{5}\nonumber\\&&-\frac{1}{10}(\gamma^{\alpha_{1}}\gamma^{\beta_{1}}
  +\frac{\gamma^{\alpha_{1}}p^{\beta_{1}}-\gamma^{\beta_{1}}p^{\alpha_{1}}}{m_{5/2}}-\frac{p^{\alpha_{1}}p^{\beta_{1}}}{m^{2}_{5/2}})
  \tilde{g}^{\alpha_{2}\beta_{2}}\nonumber\\&&-\frac{1}{10}(\gamma^{\alpha_{2}}\gamma^{\beta_{1}}
  +\frac{\gamma^{\alpha_{2}}p^{\beta_{1}}-\gamma^{\beta_{1}}p^{\alpha_{2}}}{m_{5/2}}-\frac{p^{\alpha_{2}}p^{\beta_{1}}}{m^{2}_{5/2}})
  \tilde{g}^{\alpha_{1}\beta_{2}}\nonumber\\&&-\frac{1}{10}(\gamma^{\alpha_{1}}\gamma^{\beta_{2}}
  +\frac{\gamma^{\alpha_{1}}p^{\beta_{2}}-\gamma^{\beta_{2}}p^{\alpha_{1}}}{m_{5/2}}-\frac{p^{\alpha_{1}}p^{\beta_{2}}}{m^{2}_{5/2}})
  \tilde{g}^{\alpha_{2}\beta_{1}}\nonumber\\&&-\frac{1}{10}(\gamma^{\alpha_{2}}\gamma^{\beta_{2}}
  +\frac{\gamma^{\alpha_{2}}p^{\beta_{2}}-\gamma^{\beta_{2}}p^{\alpha_{2}}}{m_{5/2}}-\frac{p^{\alpha_{2}}p^{\beta_{2}}}{m^{2}_{5/2}})
  \tilde{g}^{\alpha_{1}\beta_{1}}],
\end{eqnarray}
with $\tilde{g}^{\mu\nu}=g^{\mu\nu}-\frac{p^{\mu}p^{\nu}}{p^{2}}$.

On the other hand, the correlation function (\ref{2-point correlator}) can be calculated theoretically via OPE method at the quark-gluon level. We take the current $J_{1/2,-,\Omega_{b},1,0,\rho}(x)$ as an example to illustrate involved technologies. Inserting the interpolating current $J_{1/2,-,\Omega_{b},1,0,\rho}(x)$ (\ref{interpolating current}) into the correlation function (\ref{2-point correlator}) and contracting the relevant quark fields by Wick's theorem, we find
\begin{eqnarray}\label{OPE}
\Pi^{(OPE)}(p)=&&-4i\epsilon_{abc}\epsilon_{a^{\prime}b^{\prime}c^{\prime}}\int d^{4}x e^{ipx}\gamma_{\mu}\gamma_{5}S^{(b)}_{cc^{\prime}}(x)\gamma_{\mu^{\prime}}\gamma_{5}\nonumber\\
&&\{Tr[\gamma_{5}S^{(s)}_{bb^{\prime}}(x)\gamma_{5}C\partial^{\mu}_{x}\partial^{\mu^{\prime}}_{y}S^{(s)T}_{aa^{\prime}}(x-y)C]-
Tr[\gamma_{5}\partial^{\mu}_{x}S^{(s)}_{bb^{\prime}}(x)\gamma_{5}C\partial^{\mu^{\prime}}_{y}S^{(s)T}_{aa^{\prime}}(x-y)C]\}_{y=0}\nonumber\\
&&+4\epsilon_{abc}\epsilon_{a^{\prime}b^{\prime}c^{\prime}}\int d^{4}x e^{ipx}g_{s}A^{\mu ad}(x)\gamma_{\mu}\gamma_{5}S^{(b)}_{cc^{\prime}}(x)\gamma_{\mu^{\prime}}\gamma_{5}\nonumber\\
&&\{Tr[\gamma_{5}\partial^{\mu^{\prime}}_{y}S^{(s)}_{bb^{\prime}}(x-y)\gamma_{5}CS^{(s)T}_{da^{\prime}}(x)C]-
Tr[\gamma_{5}S^{(s)}_{bb^{\prime}}(x)\gamma_{5}C\partial^{\mu^{\prime}}_{y}S^{(s)T}_{da^{\prime}}(x-y)C]\}_{y=0}\nonumber\\
&&+\frac{\langle0|g_{s}\bar{s}\sigma\cdot Gs|0\rangle}{96}\epsilon_{abc}\epsilon_{a^{\prime}b^{\prime}c^{\prime}}\int d^{4}x e^{ipx}g_{s}\gamma_{\mu}\gamma_{5}S^{(b)}_{cc^{\prime}}(x)\gamma_{\mu^{\prime}}\gamma_{5}\nonumber\\
&&(\frac{\lambda_{n}}{2})^{ad}\{(\frac{\lambda_{n}}{2})^{da^{\prime}}x_{\nu}
Tr[\gamma_{5}\partial^{\mu^{\prime}}_{y}S^{(s)}_{bb^{\prime}}(x-y)\gamma_{5}\sigma^{\mu\nu}]-(\frac{\lambda^{n}}{2})^{bb^{\prime}}x_{\nu}
Tr[\gamma_{5}\partial^{\mu^{\prime}}_{y}S^{(s)}_{da^{\prime}}(x-y)\gamma_{5}\sigma^{\mu\nu}]\}_{y=0},
\end{eqnarray}
where $a$, $b$, $\cdots$ are color indices, $\lambda^{n},n=1,2,\cdots,8$ are Gell-Mann matrix, $A^{\mu ad}(x)=A^{n\mu}(x)(\frac{\lambda_{n}}{2})^{ad}$ is the gluon field, $g_{s}$ is the strong interaction constant and $S^{(b)}(x)$ and $S^{(s)}(x)$ are the full bottom- and strange-quark propagators, whose expressions are given in Appendix \ref{appendix1}. Inserting the expressions of full quark propagators into (\ref{OPE}) and performing involved integrals, we have
\begin{equation}\label{QCD side}
\Pi^{(OPE)}(p)=\not\!{p}(\int^{\infty}_{(m_{b}+2m_{s})^{2}}ds\frac{\rho(s)}{s-p^2}
+\frac{m^{2}_{s}\langle0|\bar{s}s|0\rangle^{2}}{12(m^{2}_{b}-p^{2})})+\mbox{other Lorentz structures},
\end{equation}
where $\rho(s)$ is the QCD spectral density
\begin{eqnarray}
\rho(s)=&&-\frac{3}{64\pi^{4}}\int^{1}_{a_{min}}da
\frac{(1-a)^{3}}{a^{2}}(m^{2}_{b}-as)^{3}+\frac{3m^{2}_{s}}{16\pi^{4}}\int^{1}_{a_{min}}da
\frac{(1-a)^{2}}{a}(m^{2}_{b}-as)^{2}\nonumber\\&&-\frac{3m_{s}\langle0|\bar{s}s|0\rangle}{4\pi^{2}}\int^{1}_{a_{min}}da
(1-a)(m^{2}_{b}-as)-\frac{m^{2}_{b}\langle0|g^{2}_{s}GG|0\rangle}{256\pi^{4}}\int^{1}_{a_{min}}da
\frac{(1-a)^{3}}{a^{2}}\nonumber\\&&-\frac{5\langle0|g^{2}_{s}GG|0\rangle}{256\pi^{4}}\int^{1}_{a_{min}}da(1-a)(m^{2}_{b}-as)
-\frac{m^{2}_{s}\langle0|g^{2}_{s}GG|0\rangle}{192\pi^{4}}(1-a_{min})^{2}\nonumber\\&&
-\frac{m_{s}\langle0|\bar{s}s|0\rangle\langle0|g^{2}_{s}GG|0\rangle}{96\pi^{2}M^{2}_{B}}(1-a_{min}),
\end{eqnarray}
with $a_{min}=m^{2}_{b}/s$, $m_{s}$ being the mass of the strange quark, $m_{b}$ being the mass of the bottom quark and $M^{2}_{B}$ being the Borel parameter introduced as making Borel transform in the next step.

Finally, we match the phenomenological side (\ref{hadronic side}) and the QCD representation (\ref{QCD side}) for the Lorentz structure $\not\!{p}$,
\begin{eqnarray}
\frac{f^{2}_{1/2,-,\Omega_{b},1,0,\rho}}{m^{2}_{1/2,-,\Omega_{b},1,0,\rho}-p^{2}}+\mbox{higher resonances}=\int^{\infty}_{(m_{b}+2m_{s})^{2}}ds\frac{\rho(s)}{s-p^2}
+\frac{m^{2}_{s}\langle0|\bar{s}s|0\rangle^{2}}{12(m^{2}_{b}-p^{2})},
\end{eqnarray}
According to the quark-hadron duality, the higher resonances can be approximated by the QCD spectral density above some effective threshold $s^{1/2,-,\Omega_{b},1,0,\rho}_{0}$,
\begin{equation}
\frac{f^{2}_{1/2,-,\Omega_{b},1,0,\rho}}{m^{2}_{1/2,-,\Omega_{b},1,0,\rho}-p^{2}}
+\int^{\infty}_{s^{1/2,-,\Omega_{b},1,0,\rho}_{0}}ds\frac{\rho(s)}{s-p^2}=\int^{\infty}_{(m_{b}+2m_{s})^{2}}ds\frac{\rho(s)}{s-p^2}
+\frac{m^{2}_{s}\langle0|\bar{s}s|0\rangle^{2}}{12(m^{2}_{b}-p^{2})}.
\end{equation}
Subtracting the contributions of the excited and continuum states, one gets
\begin{equation}
\frac{f^{2}_{1/2,-,\Omega_{b},1,0,\rho}}{m^{2}_{1/2,-,\Omega_{b},1,0,\rho}-p^{2}}
=\int^{s^{1/2,-,\Omega_{b},1,0,\rho}_{0}}_{(m_{b}+2m_{s})^{2}}ds\frac{\rho(s)}{s-p^2}
+\frac{m^{2}_{s}\langle0|\bar{s}s|0\rangle^{2}}{12(m^{2}_{b}-p^{2})},
\end{equation}
To improve the convergence of the OPE series and suppress the contributions from the excited and continuum states, it is necessary to make a Borel transform. As a result, we have
\begin{equation}\label{mass sum rule1}
f^{2}_{1/2,-,\Omega_{b},1,0,\rho}e^{-m^{2}_{1/2,-,\Omega_{b},1,0,\rho}/M^{2}_{B}}
=\int^{s^{1/2,-,\Omega_{b},1,0,\rho}_{0}}_{(m_{b}+2m_{s})^{2}}ds\rho(s)e^{-s/M^{2}_{B}}
+\frac{m^{2}_{s}\langle0|\bar{s}s|0\rangle^{2}}{12}e^{-m^{2}_{b}/M^{2}_{B}},
\end{equation}
where $M^{2}_{B}$ is the Borel parameter. Applying the operator $-\frac{d}{d(1/M^{2}_{B})}$ to (\ref{mass sum rule1}) and dividing the resulting equation with (\ref{mass sum rule1}), we obtain the mass sum rule
\begin{equation}\label{mass sum rule}
m^{2}_{1/2,-,\Omega_{b},1,0,\rho}=\frac{-\frac{d}{d(1/M^{2}_{B})}(\int^{s^{1/2,-,\Omega_{b},1,0,\rho}_{0}}_{(m_{b}+2m_{s})^{2}}ds\rho(s)e^{-s/M^{2}_{B}}
+\frac{m^{2}_{s}\langle0|\bar{s}s|0\rangle^{2}}{12}e^{-m^{2}_{b}/M^{2}_{B}})}{
\int^{s^{1/2,-,\Omega_{b},1,0,\rho}_{0}}_{(m_{b}+2m_{s})^{2}}ds\rho(s)e^{-s/M^{2}_{B}}
+\frac{m^{2}_{s}\langle0|\bar{s}s|0\rangle^{2}}{12}e^{-m^{2}_{b}/M^{2}_{B}}}.
\end{equation}
In Sec.\ref{sec3}, we will numerically analyze (\ref{mass sum rule}) and (\ref{mass sum rule1}) and estimate the values of the mass $m_{1/2,-,\Omega_{b},1,0,\rho}$ and the pole residue $f_{1/2,-,\Omega_{b},1,0,\rho}$.

For other interpolating currents, we do the same analysis and the corresponding OPE results are given in Appendix \ref{appendix2}.

\section{Numerical analysis}\label{sec3}

The sum rule (\ref{mass sum rule}) contains some parameters, various condensates and quark masses, whose values are presented in Table \ref{input parameters}. The values of $m_{b}$ and $m_{s}$ are the $\overline{MS}$ values. Besides these parameters, we should determine the working intervals of the threshold parameter $s^{j,P,\Omega_{b},j_{l},s_{l},\rho/\lambda}_{0}$ and the Borel mass $M^{2}_{B}$ in which the masses and pole residues is stable. We take the continuum threshold to be around $m_{j,P,\Omega_{b},j_{l},s_{l},\rho/\lambda}+(0.7\pm0.1)\mbox{GeV}$, while the Borel parameter is determined by demanding that both the contributions of the higher states and continuum are sufficiently suppressed and the contributions coming from higher dimensional operators are small.
\begin{table}[htb]
\caption{Some input parameters needed in the calculations.}\label{input parameters}
\begin{tabular}{|c|c|}
  \hline
  Parameter      &   Value    \\
  \hline
  {$\langle\bar{s}s\rangle$}  &      $(0.8\pm0.1)\langle\bar{q}q\rangle$                     \\
  {$\langle\bar{q}q\rangle$}  &      $-(0.24\pm0.01)^{3}\mbox{GeV}^{3}$                     \\
  {$\langle g_{s}\bar{s}\sigma Gs\rangle$} & $(0.8\pm0.1)\langle\bar{s}s\rangle \mbox{GeV}^{2}$ \\
  {$\langle g^{2}_{s}GG\rangle$}    &     $0.88\pm0.25\mbox{GeV}^{4}$                \\
  {$m_{b}$}  &    $(4.18\pm0.03)\mbox{GeV}$\cite{M.Tanabashi}                   \\
  {$m_{s}$}  &   $(0.095\pm0.005)\mbox{GeV}$\cite{M.Tanabashi}     \\
  \hline
\end{tabular}
\end{table}

We define two quantities, the ratio of the pole contribution to the total contribution (Pole Contribution abbreviated as PC) and the ratio of the highest dimensional term in the OPE series to the total OPE series (Convergence abbreviated as CVG), as followings,
\begin{eqnarray}
&&\mbox{PC}\equiv\frac{\int^{s^{j,P,\Omega_{b},j_{l},s_{l},\rho/\lambda}_{0}}_{(m_{b}+2m_{s})^{2}}ds\rho(s)e^{-\frac{s}{M^{2}_{B}}}}
{\int^{\infty}_{(m_{b}+2m_{s})^{2}}ds\rho(s)e^{-\frac{s}{M^{2}_{B}}}},
\nonumber\\&&\mbox{CVG}\equiv\frac{\int^{s^{j,P,\Omega_{b},j_{l},s_{l},\rho/\lambda}_{0}}_{(m_{b}+2m_{s})^{2}}ds\rho^{(d=7)}(s)e^{-\frac{s}{M^{2}_{B}}}}
{\int^{s^{j,P,\Omega_{b},j_{l},s_{l},\rho/\lambda}_{0}}_{(m_{b}+2m_{s})^{2}}ds\rho(s)e^{-\frac{s}{M^{2}_{B}}}},
\end{eqnarray}
where $\rho^{(d=7)}(s)$ is the terms proportional to $\langle0|\bar{s}s|0\rangle\langle0|g^{2}_{s}GG|0\rangle$ in spectral density.

\begin{figure}[htb]
\subfigure[]{
\includegraphics[width=7cm]{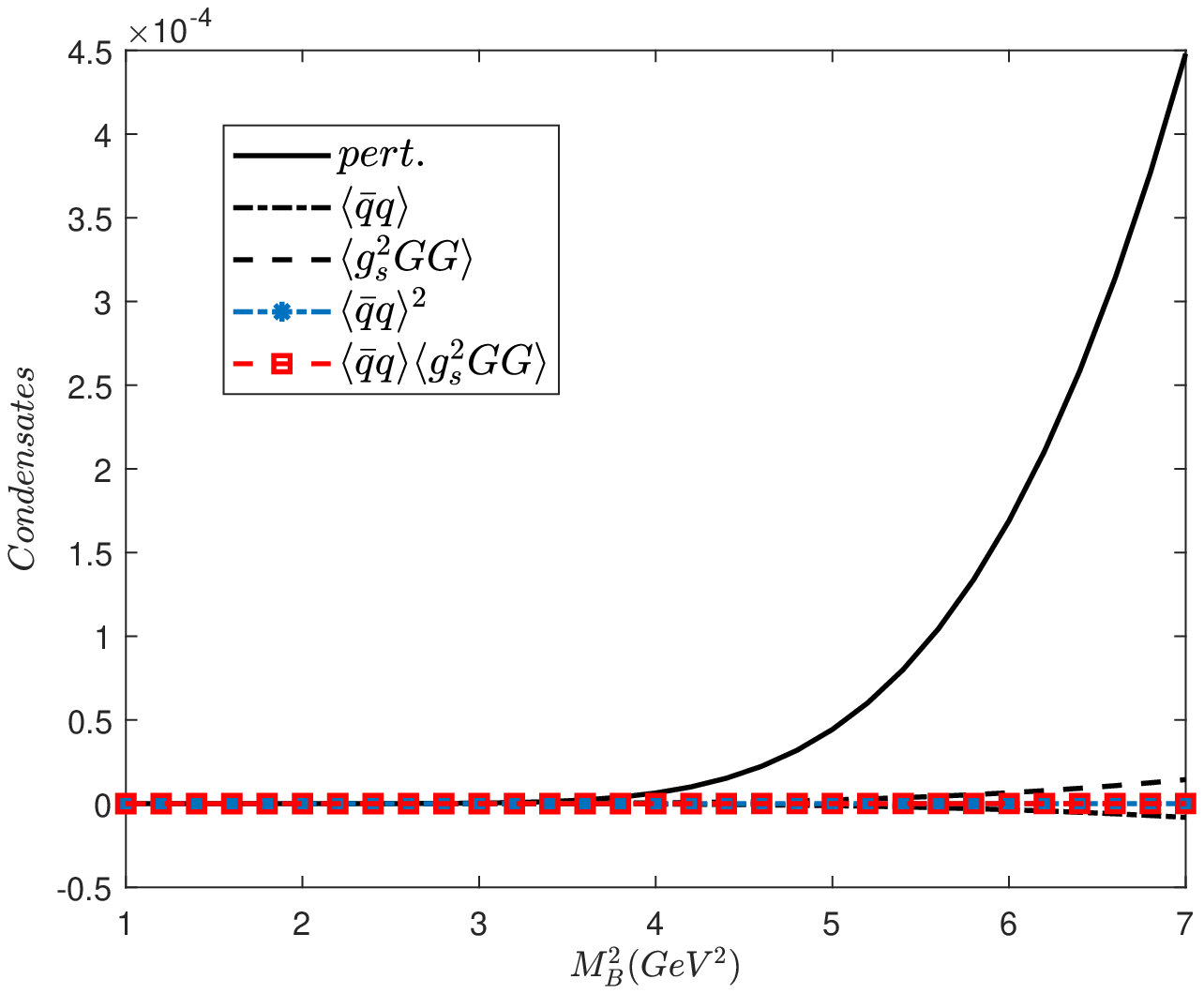}}
\subfigure[]{
\includegraphics[width=7cm]{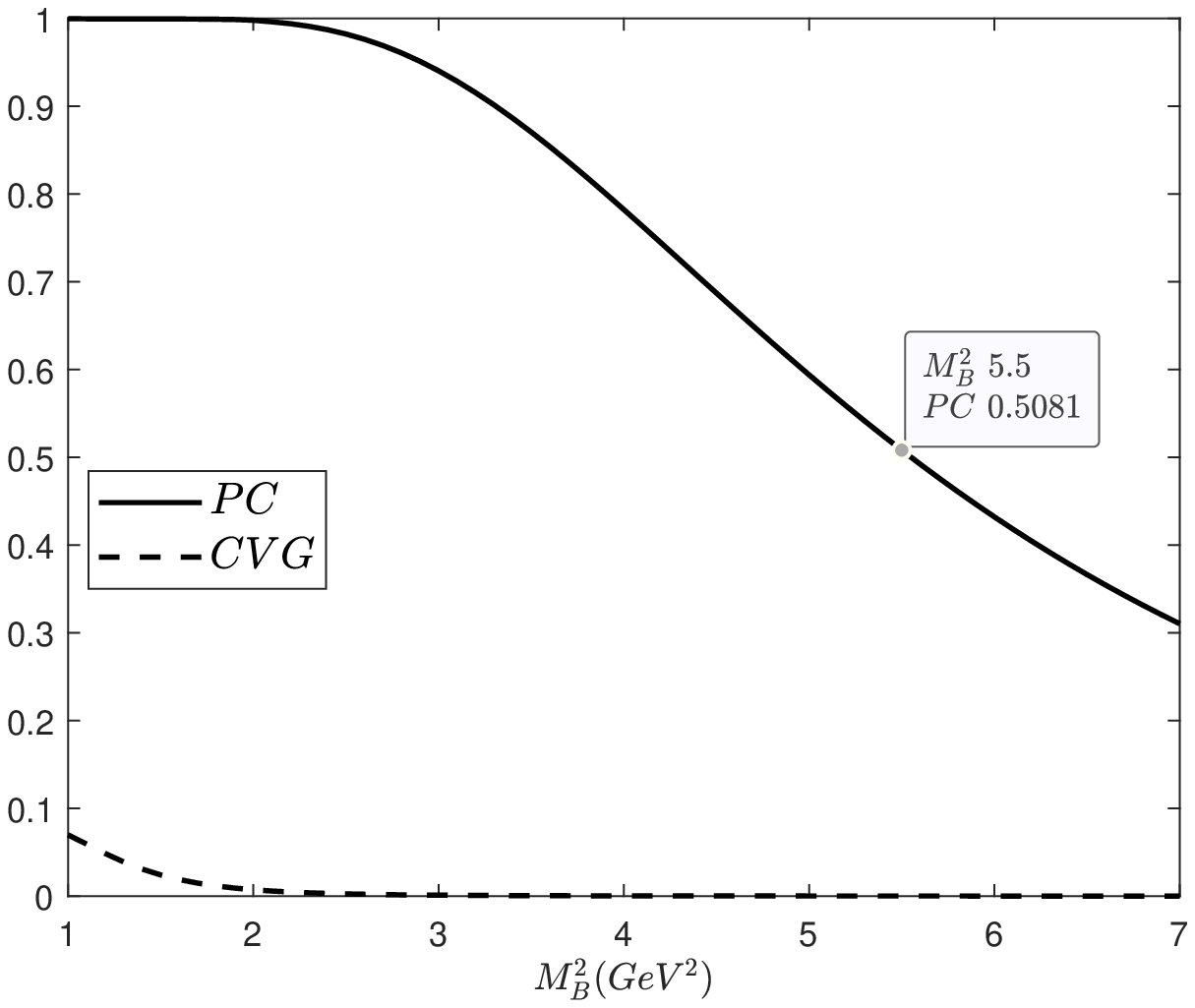}}
\subfigure[]{
\includegraphics[width=7cm]{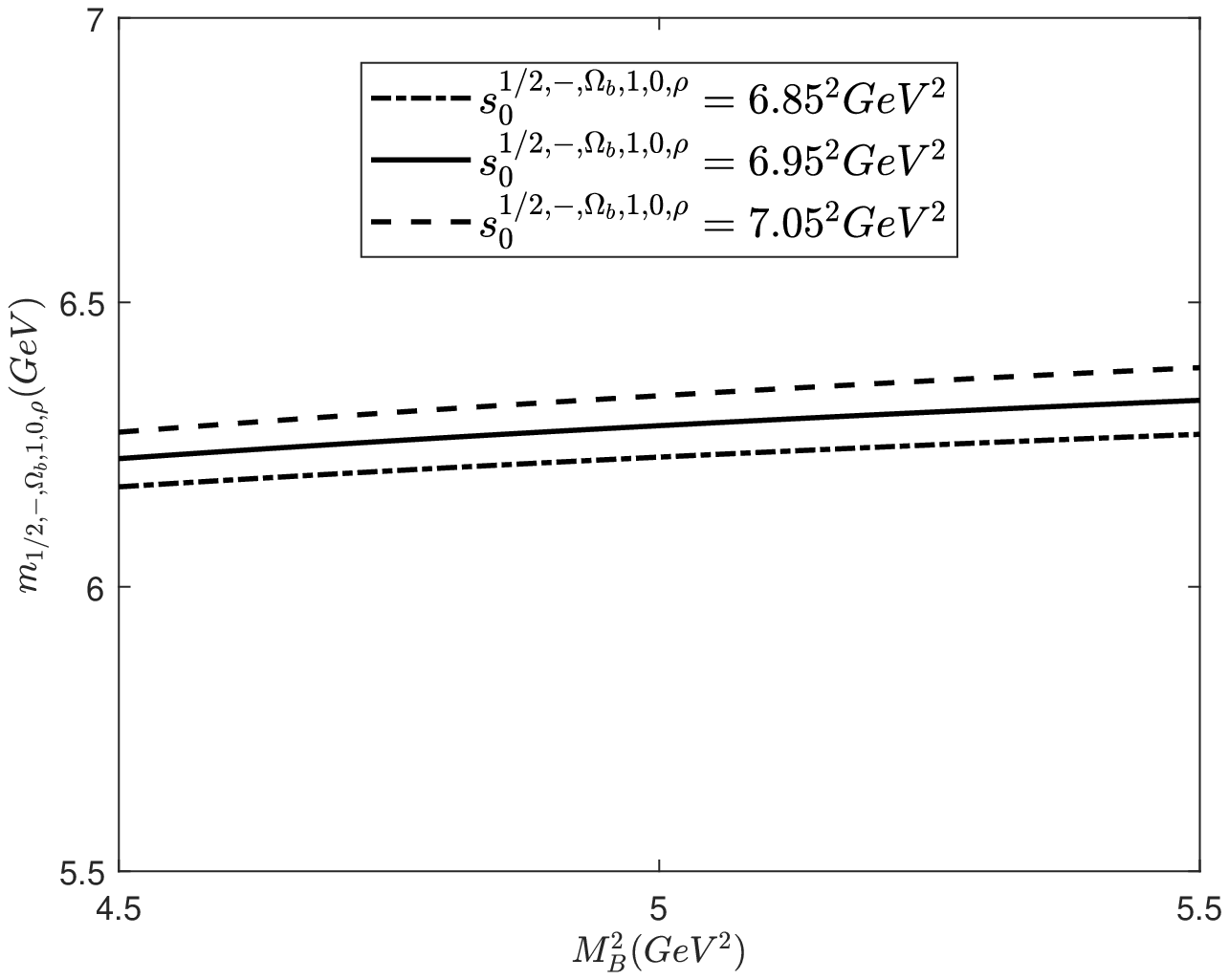}}
\subfigure[]{
\includegraphics[width=7cm]{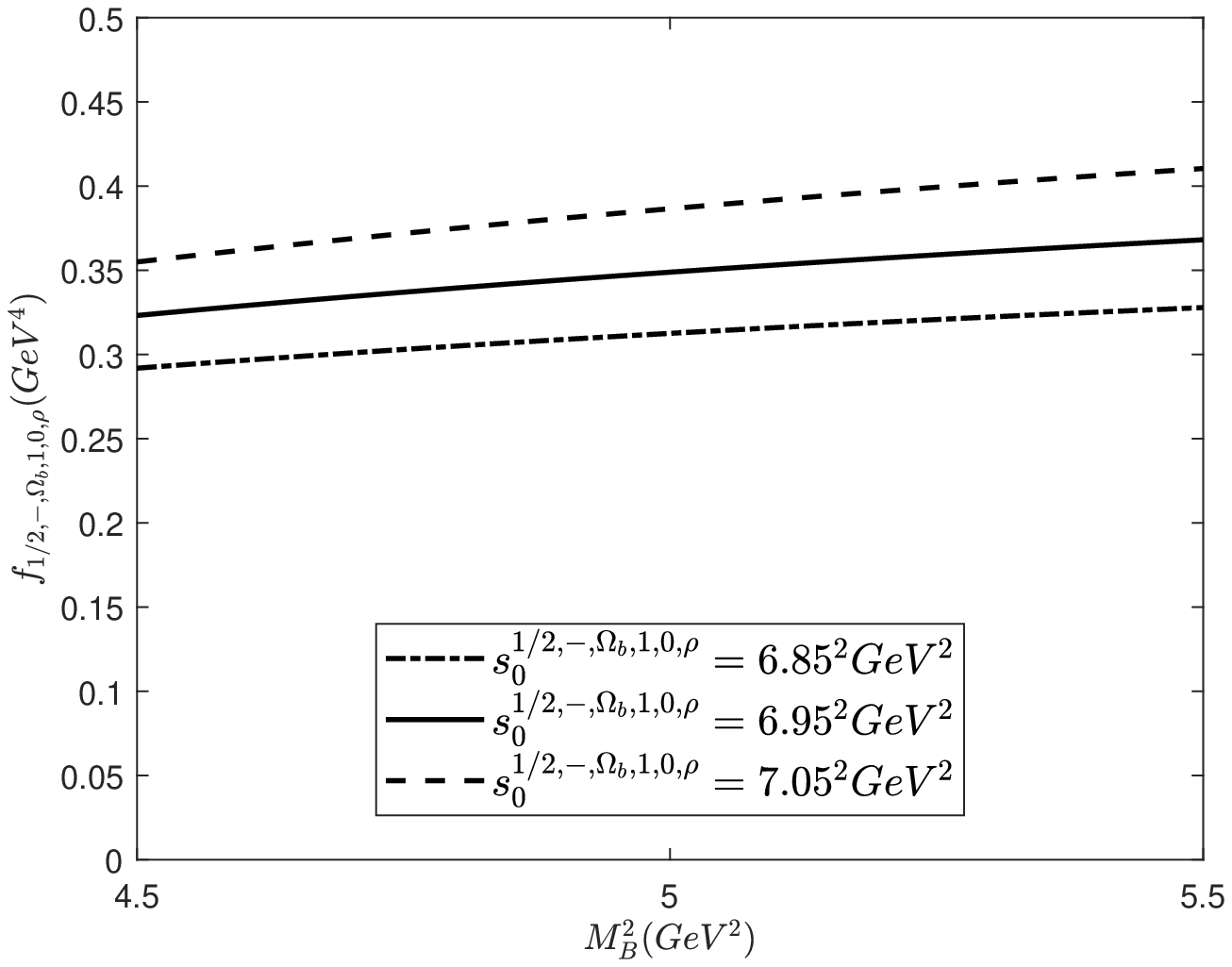}}
\caption{For the interpolating current $J_{1/2,-,\Omega_{b},1,0,\rho}(x)$: (a) denotes the various condensate contributions as functions of $M^{2}_{B}$ with $s^{1/2,-,\Omega_{b},1,0,\rho}_{0}=6.95^{2}\mbox{GeV}^{2}$; (b) represents PC and CVG varying with $M^{2}_{B}$ at $s^{1/2,-,\Omega_{b},1,0,\rho}_{0}=6.95^{2}\mbox{GeV}^{2}$; (c) and (d) depict the dependence of the mass and the pole residue on $M^{2}_{B}$ with three different values of $s^{1/2,-,\Omega_{b},1,0,\rho}_{0}$, respectively.}\label{result1}
\end{figure}

For the current $J_{1/2,-,\Omega_{b},1,0,\rho}(x)$, the numerical results are shown in Fig.\ref{result1}. In Fig.\ref{result1}(a), we compare the various condensate contributions as functions of $M^{2}_{B}$ with $s^{1/2,-,\Omega_{b},1,0,\rho}_{0}=6.95^{2}\mbox{GeV}^{2}$. From it one can see that the OPE has good convergence. Fig.\ref{result1}(b) shows PC and CVG varying with $M^{2}_{B}$ at $s^{1/2,-,\Omega_{b},1,0,\rho}_{0}=6.95^{2}\mbox{GeV}^{2}$. The figure shows that the requirement $\mbox{PC}\geq50\%$ gives $M^{2}_{B}\leq5.5\mbox{GeV}^{2}$. The dependence of the mass $m_{1/2,-,\Omega_{b},1,0,\rho}$ and the pole residue $f_{1/2,-,\Omega_{b},1,0,\rho}$ on the Borel parameter $M^{2}_{B}$ are depicted in Fig.\ref{result1}(c) and (d) at three different values of $s^{1/2,-,\Omega_{b},1,0,\rho}_{0}$, respectively. It is obvious that the mass and the pole residue are stable in the interval $4.5\mbox{GeV}^{2}\leq M^{2}_{B}\leq5.5\mbox{GeV}^{2}$. The mass and the pole residue are estimated to be $m_{1/2,-,\Omega_{b},1,0,\rho}=(6.28^{+0.11}_{-0.10})\mbox{GeV}$ and $f_{1/2,-,\Omega_{b},1,0,\rho}=(0.35\pm0.06)\mbox{GeV}^{4}$, respectively.

\begin{figure}[htb]
\subfigure[]{
\includegraphics[width=7cm]{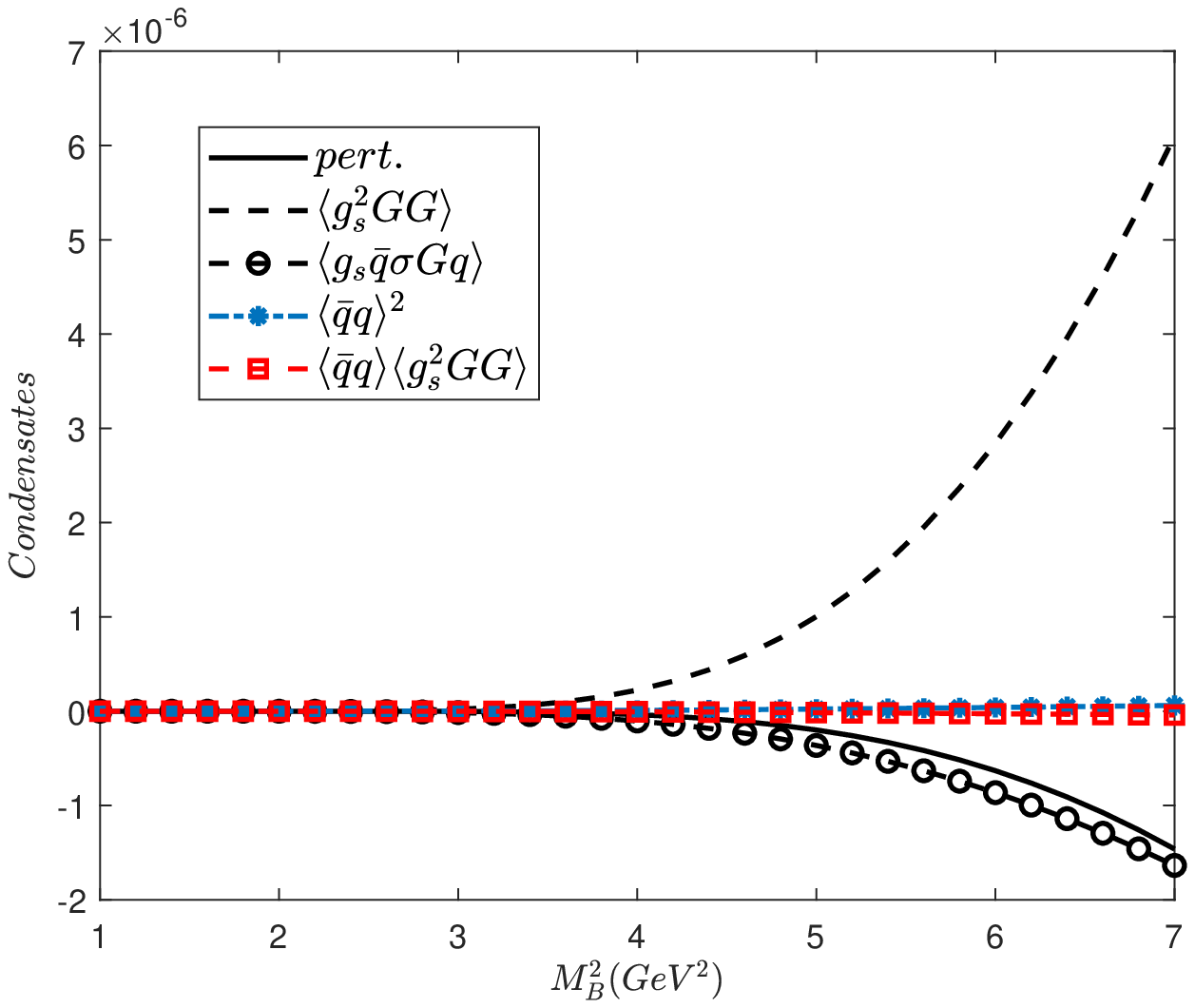}}
\subfigure[]{
\includegraphics[width=7cm]{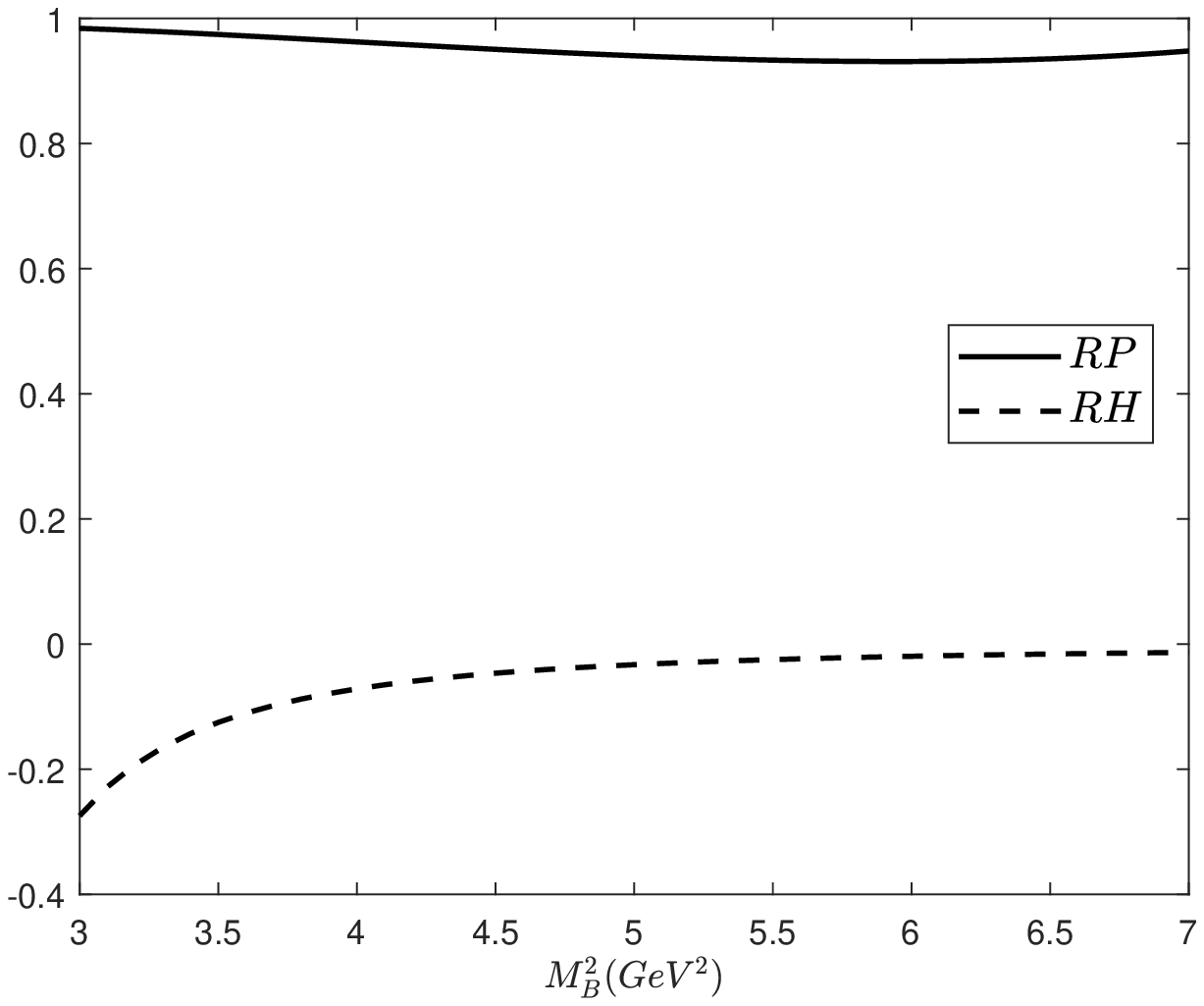}}
\subfigure[]{
\includegraphics[width=7cm]{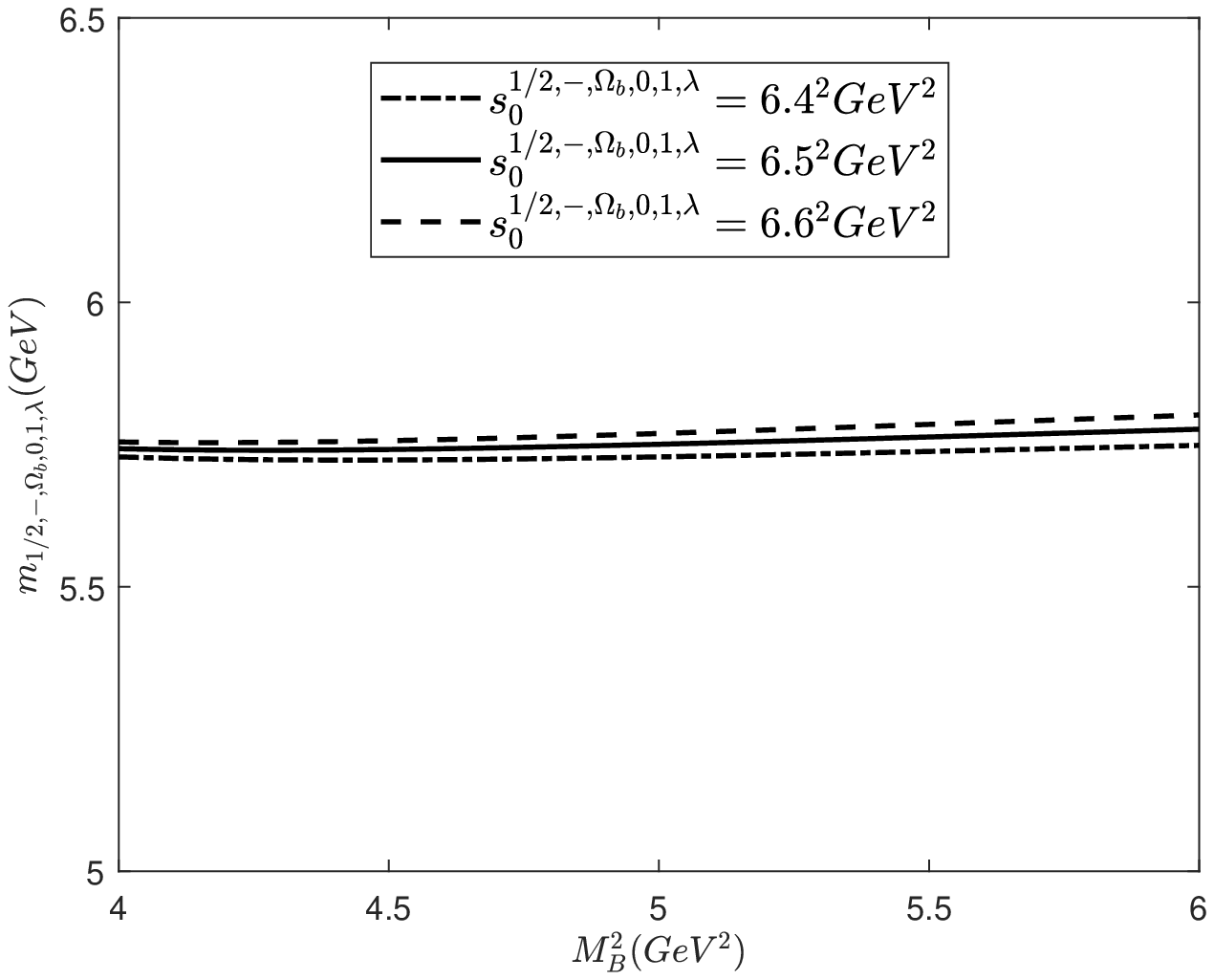}}
\subfigure[]{
\includegraphics[width=7cm]{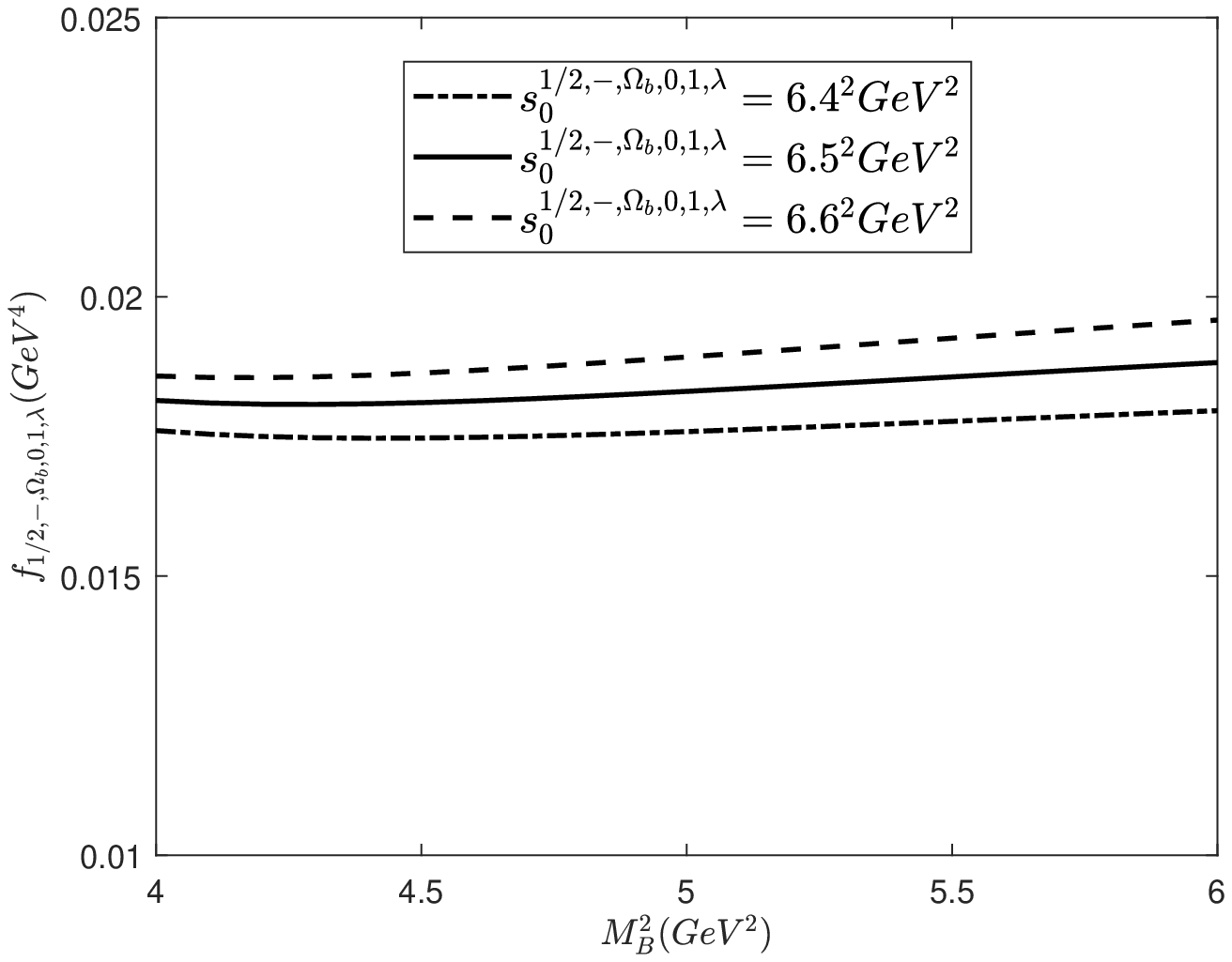}}
\caption{For the interpolating current $J_{1/2,-,\Omega_{b},0,1,\lambda}(x)$: (a) denotes the various condensate contributions as functions of $M^{2}_{B}$ with $s^{1/2,-,\Omega_{b},0,1,\lambda}_{0}=6.5^{2}\mbox{GeV}^{2}$; (b) represents $RP$ and $RH$ varying with $M^{2}_{B}$ at $s^{1/2,-,\Omega_{b},0,1,\lambda}_{0}=6.5^{2}\mbox{GeV}^{2}$; (c) and (d) depict the dependence of the mass and the pole residue on $M^{2}_{B}$ with three different values of $s^{1/2,-,\Omega_{b},0,1,\lambda}_{0}$, respectively.}\label{result3}
\end{figure}

For other interpolating currents, the same analysis can be done. We summarize our results in Table \ref{results} and compare the obtained masses with the results in Ref.\cite{omegab2} estimated by QCD sum rule method in the framework of heavy quark effective theory. We can see that they are agreement with each other within the inherent uncertainties of the QCD sum rule method except for the multiplet [$\Omega_{b}$, 0, 1, $\lambda$]. We should give some arguments about the result of the interpolating current $J_{1/2,-,\Omega_{b},0,1,\lambda}(x)$ shown in Fig.\ref{result3}. From Eqs.(\ref{argument1}) and (\ref{argument2}), we can see that all terms of the OPE series are proportional to the strange quark mass $m_{s}$ or $m^{2}_{s}$ except for the second term in (\ref{argument2}). As a result, the gluon-condensate term is much larger than other terms and OPE is invalid in this case. Moreover, the corresponding mass and pole residue are much lower than others. All in all, our model can not give reasonable results in this case.
\begin{table}[htb]
\caption{The masses and pole residues of the P-wave excited $\Omega_{b}$ states.}\label{results}
\begin{tabular}{|c|c|c|c|c|}
\hline
\multirow{2}{*}{Multiples} & \multirow{2}{*}{Baryons($j^{P}$)} & \multicolumn{2}{|c|}{Masses($\mbox{GeV}$)} & \multirow{2}{*}{Pole residues($\mbox{GeV}^{4}$)} \\
\cline{3-4} & & {This work} & {Ref.\cite{omegab2}} & \\
\hline
\multirow{2}{*}{[$\Omega_{b}$, 1, 0, $\rho$]} & {$\Omega_{b}$($\frac{1}{2}^{-}$)} & {$6.28^{+0.11}_{-0.10}$} & {$6.32^{+0.12}_{-0.10}$} & {$0.35\pm0.06$} \\
\cline{2-5} & {$\Omega_{b}$($\frac{3}{2}^{-}$)} & {$6.31^{+0.10}_{-0.11}$} & {$6.32^{+0.12}_{-0.10}$} & {$0.19\pm0.03$} \\
\hline
{[$\Omega_{b}$, 0, 1, $\lambda$]} & {$\Omega_{b}$($\frac{1}{2}^{-}$)} & {$5.75^{+0.05}_{-0.02}$} & {$6.34\pm0.11$} & {$0.0183^{+0.0013}_{-0.0007}$} \\
\hline
\multirow{2}{*}{[$\Omega_{b}$, 1, 1, $\lambda$]}  & {$\Omega_{b}$($\frac{1}{2}^{-}$)} & {$6.33^{+0.10}_{-0.11}$} & {$6.34^{+0.09}_{-0.08}$} & {$0.62\pm0.10$} \\
\cline{2-5} & {$\Omega_{b}$($\frac{3}{2}^{-}$)} & {$6.37^{+0.10}_{-0.11}$} & {$6.34^{+0.09}_{-0.08}$} & {$0.36^{+0.06}_{-0.05}$} \\
\hline
\multirow{2}{*}{[$\Omega_{b}$, 2, 1, $\lambda$]}  & {$\Omega_{b}$($\frac{3}{2}^{-}$)} & {$6.34^{+0.09}_{-0.10}$} & {$6.35^{+0.13}_{-0.11}$} & {$0.71\pm0.11$} \\
\cline{2-5} & {$\Omega_{b}$($\frac{5}{2}^{-}$)} & {$6.54^{+0.07}_{-0.08}$} & {$6.36^{+0.13}_{-0.11}$} & {$0.15\pm0.02$}  \\
\hline
\end{tabular}
\end{table}

\section{Conclusion}\label{sec4}

In this paper, we consider all P-wave $\Omega_{b}$ states represented by interpolating currents with a derivative and calculate the corresponding masses and pole residues with the method of QCD sum rule. The results are listed in Table \ref{results}. Due to the large uncertainties in our calculation compared with the small difference in the masses of the excited $\Omega_{b}$ states observed by the LHCb collaboration, it is necessary to study other properties of the P-wave $\Omega_{b}$ states represented by the interpolating currents investigated in the present work in order to have a better understanding about the four excited $\Omega_{b}$ states observed by the LHCb collaboration. For example, we could study their decay widths. Our results in this paper are necessary input parameters when studying their decay widths by QCD sum rule method or light-cone sum rule method.

\acknowledgments  One of the authors, Yong-Jiang Xu, thanks Hua-Xing Chen for useful discussion on the construction of interpolating currents. This work was supported by the National
Natural Science Foundation of China under Contract No.11675263.

\begin{appendix}
\section{The quark propagators}\label{appendix1}
The full quark propagators are
\begin{eqnarray}
 S^{q}_{ij}(x)=&&\frac{i \not\!{x}}{2\pi^{2}x^4}\delta_{ij}-\frac{m_{q}}{4\pi^2x^2}\delta_{ij}-\frac{\langle\bar{q}q\rangle}{12}\delta_{ij}
 +i\frac{\langle\bar{q}q\rangle}{48}m_{q}\not\!{x}\delta_{ij}-\frac{x^2}{192}\langle g_{s}\bar{q}\sigma Gq\rangle \delta_{ij}\nonumber\\
 &&+i\frac{x^2\not\!{x}}{1152}m_{q}\langle g_{s}\bar{q}\sigma Gq\rangle \delta_{ij}-i\frac{g_{s}t^{a}_{ij}G^{a}_{\mu\nu}}{32\pi^2x^2}(\not\!{x}\sigma^{\mu\nu}+\sigma^{\mu\nu}\not\!{x})+\cdots
 \end{eqnarray}
 for light quark, and
 \begin{eqnarray}
 S^{Q}_{ij}(x)=i\int\frac{d^{4}k}{(2\pi)^4}e^{-ikx}&&[\frac{\not\!{k}+m_{Q}}{k^2-m^{2}_{Q}}\delta_{ij}
 -\frac{g_{s}t^{a}_{ij}G^{a}_{\mu\nu}}{4}\frac{\sigma^{\mu\nu}(\not\!{k}+m_{Q})+(\not\!{k}+m_{Q})\sigma^{\mu\nu}}
 {(k^2-m^{2}_{Q})^{2}}\nonumber\\
 &&+\frac{\langle g^{2}_{s}GG\rangle}{12}\delta_{ij}m_{Q}\frac{k^2+m_{Q}\not\!{k}}{(k^2-m^{2}_{Q})^{4}}+\cdots]
 \end{eqnarray}
 for heavy quark. In these expressions, $t^{a}=\frac{\lambda^{a}}{2}$ and $\lambda^{a}$ are the Gell-Mann matrices, $g_{s}$ is the strong interaction coupling constant, and $i, j$ are color indices.

\section{The spectral densities}\label{appendix2}
 We choose the Lorentz structure $\not\!{p}$, $\not\!{p}g^{\alpha\beta}$ and $\not\!{p}g^{\alpha_{1}\alpha_{2}}g^{\beta_{1}\beta_{2}}$ to obtain the sum rules for spin-$\frac{1}{2}$, spin-$\frac{3}{2}$ and spin-$\frac{5}{2}$ baryons, respectively. In this appendix, we will give the corresponding OPE results.

For the interpolating current $J^{\alpha}_{3/2,-,\Omega_{b},1,0,\rho}(x)$,
\begin{equation}
\Pi^{(OPE)\alpha\beta}(p)=\not\!{p}g^{\alpha\beta}(\int^{\infty}_{(m_{b}+2m_{s})^{2}}ds\frac{\rho(s)}{s-p^2}
-\frac{m^{2}_{s}\langle0|\bar{s}s|0\rangle^{2}}{24(m^{2}_{b}-p^{2})})+\mbox{other Lorentz structures},
\end{equation}
where $\rho(s)$ is the QCD spectral density,
\begin{eqnarray}
\rho(s)=&&\frac{1}{384\pi^{4}}\int^{1}_{a_{min}}da
\frac{(1-a)^{3}(4+a)}{a^{2}}(m^{2}_{b}-as)^{3}-\frac{m^{2}_{s}}{64\pi^{4}}\int^{1}_{a_{min}}da
\frac{(1-a)^{2}(2+a)}{a}(m^{2}_{b}-as)^{2}\nonumber\\&&+\frac{m_{s}\langle0|\bar{s}s|0\rangle}{8\pi^{2}}\int^{1}_{a_{min}}da
a(1-a)(m^{2}_{b}-as)+\frac{m^{2}_{b}\langle0|g^{2}_{s}GG|0\rangle}{4608\pi^{4}}\int^{1}_{a_{min}}da
\frac{(1-a)^{3}(4+a)}{a^{2}}\nonumber\\&&+\frac{\langle0|g^{2}_{s}GG|0\rangle}{512\pi^{4}}\int^{1}_{a_{min}}da(1-a)(2+a)(m^{2}_{b}-as)
+\frac{m^{2}_{s}\langle0|g^{2}_{s}GG|0\rangle}{2304\pi^{4}}(1-a_{min})^{2}(2+a_{min})\nonumber\\&&
+\frac{m_{s}\langle0|g_{s}\bar{s}\sigma\cdot Gs|0\rangle}{24\pi^{2}}\int^{1}_{a_{min}}daa(2-a)+\frac{m_{s}\langle0|\bar{s}s|0\rangle\langle0|g^{2}_{s}GG|0\rangle}{576\pi^{2}M^{2}_{B}}a_{min}(1-a_{min}).
\end{eqnarray}

For the interpolating current $J_{1/2,-,\Omega_{b},0,1,\lambda}(x)$,
\begin{eqnarray}\label{argument1}
\Pi^{(OPE)}(p)=&&\not\!{p}(\int^{\infty}_{(m_{b}+2m_{s})^{2}}ds\frac{\rho(s)}{s-p^2}
+\frac{2m^{2}_{s}\langle0|\bar{s}s|0\rangle^{2}}{3(m^{2}_{b}-p^{2})}\nonumber\\
&&+\frac{m_{s}\langle0|\bar{s}s|0\rangle\langle0|g^{2}_{s}GG|0\rangle}{192\pi^{2}(m^{2}_{b}-p^{2})})+\mbox{other Lorentz structures},
\end{eqnarray}
where $\rho(s)$ is the QCD spectral density,
\begin{eqnarray}\label{argument2}
\rho(s)=&&-\frac{3m^{2}_{s}}{32\pi^{4}}\int^{1}_{a_{min}}da
\frac{(1-a)^{2}}{a}(m^{2}_{b}-as)^{2}-\frac{\langle0|g^{2}_{s}GG|0\rangle}{128\pi^{4}}\int^{1}_{a_{min}}da(1-a)(m^{2}_{b}-as)
\nonumber\\&&+\frac{m^{2}_{s}\langle0|g^{2}_{s}GG|0\rangle}{384\pi^{4}}(1-a_{min})^{2}
+\frac{3m_{s}\langle0|g_{s}\bar{s}\sigma\cdot Gs|0\rangle}{16\pi^{2}}\int^{1}_{a_{min}}daa.
\end{eqnarray}

For the interpolating current $J_{1/2,-,\Omega_{b},1,1,\lambda}(x)$,
\begin{equation}
\Pi^{(OPE)}(p)=\not\!{p}(\int^{\infty}_{(m_{b}+2m_{s})^{2}}ds\frac{\rho(s)}{s-p^2}
+\frac{m_{s}\langle0|\bar{s}s|0\rangle\langle0|g^{2}_{s}GG|0\rangle}{192\pi^{2}(m^{2}_{b}-p^{2})})+\mbox{other Lorentz structures},
\end{equation}
where $\rho(s)$ is the QCD spectral density,
\begin{eqnarray}
\rho(s)=&&-\frac{1}{8\pi^{4}}\int^{1}_{a_{min}}da
\frac{(1-a)^{3}}{a^{2}}(m^{2}_{b}-as)^{3}+\frac{27m^{2}_{s}}{32\pi^{4}}\int^{1}_{a_{min}}da
\frac{(1-a)^{2}}{a}(m^{2}_{b}-as)^{2}\nonumber\\&&+\frac{3m_{s}\langle0|\bar{s}s|0\rangle}{\pi^{2}}\int^{1}_{a_{min}}da
(1-a)(m^{2}_{b}-as)-\frac{m^{2}_{b}\langle0|g^{2}_{s}GG|0\rangle}{96\pi^{4}}\int^{1}_{a_{min}}da
\frac{(1-a)^{3}}{a^{2}}\nonumber\\&&-\frac{3\langle0|g^{2}_{s}GG|0\rangle}{128\pi^{4}}\int^{1}_{a_{min}}da\frac{(1-a)^{2}}{a}(m^{2}_{b}-as)
+\frac{\langle0|g^{2}_{s}GG|0\rangle}{128\pi^{4}}\int^{1}_{a_{min}}da(1-a)(m^{2}_{b}-as)\nonumber\\&&
-\frac{3m^{2}_{s}\langle0|g^{2}_{s}GG|0\rangle}{128\pi^{4}}(1-a_{min})^{2}-\frac{m_{s}\langle0|g_{s}\bar{s}\sigma\cdot Gs|0\rangle}{16\pi^{2}}\int^{1}_{a_{min}}da(4-7a)\nonumber\\&&
+\frac{m_{s}\langle0|\bar{s}s|0\rangle\langle0|g^{2}_{s}GG|0\rangle}{24\pi^{2}M^{2}_{B}}(1-a_{min})
-\frac{m_{s}\langle0|\bar{s}s|0\rangle\langle0|g^{2}_{s}GG|0\rangle}{96\pi^{2}s}a_{min}.
\end{eqnarray}

For the interpolating current $J^{\alpha}_{3/2,-,\Omega_{b},1,1,\lambda}(x)$,
\begin{equation}
\Pi^{(OPE)\alpha\beta}(p)=\not\!{p}g^{\alpha\beta}(\int^{\infty}_{(m_{b}+2m_{s})^{2}}ds\frac{\rho(s)}{s-p^2}
-\frac{m_{s}\langle0|\bar{s}s|0\rangle\langle0|g^{2}_{s}GG|0\rangle}{576\pi^{2}(m^{2}_{b}-p^{2})})+\mbox{other Lorentz structures},
\end{equation}
where $\rho(s)$ is the QCD spectral density,
\begin{eqnarray}
\rho(s)=&&\frac{1}{96\pi^{4}}\int^{1}_{a_{min}}da
\frac{(1-a)^{3}(3+a)}{a^{2}}(m^{2}_{b}-as)^{3}-\frac{3m^{2}_{s}}{32\pi^{4}}\int^{1}_{a_{min}}da
\frac{(1-a)^{2}(2+a)}{a}(m^{2}_{b}-as)^{2}\nonumber\\&&-\frac{m_{s}\langle0|\bar{s}s|0\rangle}{2\pi^{2}}\int^{1}_{a_{min}}da
(1-a)(1+a)(m^{2}_{b}-as)+\frac{m^{2}_{b}\langle0|g^{2}_{s}GG|0\rangle}{1152\pi^{4}}\int^{1}_{a_{min}}da
\frac{(1-a)^{3}(3+a)}{a^{2}}\nonumber\\&&-\frac{\langle0|g^{2}_{s}GG|0\rangle}{768\pi^{4}}\int^{1}_{a_{min}}da\frac{(1-a)^{2}(4-a)}{a}(m^{2}_{b}-as)
-\frac{\langle0|g^{2}_{s}GG|0\rangle}{768\pi^{4}}\int^{1}_{a_{min}}da(1-a)(1+a)(m^{2}_{b}-as)\nonumber\\&&
+\frac{m^{2}_{s}\langle0|g^{2}_{s}GG|0\rangle}{384\pi^{4}}(1-a_{min})^{2}(2+a_{min})-\frac{m_{s}\langle0|g_{s}\bar{s}\sigma\cdot Gs|0\rangle}{48\pi^{2}}\int^{1}_{a_{min}}da(3-4a+4a^{2})\nonumber\\&&
-\frac{m_{s}\langle0|\bar{s}s|0\rangle\langle0|g^{2}_{s}GG|0\rangle}{144\pi^{2}M^{2}_{B}}(1-a_{min})(1+a_{min})
-\frac{m_{s}\langle0|\bar{s}s|0\rangle\langle0|g^{2}_{s}GG|0\rangle}{288\pi^{2}s}a_{min}(1-a_{min}).
\end{eqnarray}

For the interpolating current $J^{\alpha}_{3/2,-,\Omega_{b},2,1,\lambda}(x)$,
\begin{equation}
\Pi^{(OPE)\alpha\beta}(p)=\not\!{p}g^{\alpha\beta}(\int^{\infty}_{(m_{b}+2m_{s})^{2}}ds\frac{\rho(s)}{s-p^2}
+\frac{5m_{s}\langle0|\bar{s}s|0\rangle\langle0|g^{2}_{s}GG|0\rangle}{576\pi^{2}(m^{2}_{b}-p^{2})})+\mbox{other Lorentz structures},
\end{equation}
where $\rho(s)$ is the QCD spectral density,
\begin{eqnarray}
\rho(s)=&&\frac{1}{96\pi^{4}}\int^{1}_{a_{min}}da
\frac{(1-a)^{3}(7+13a)}{a^{2}}(m^{2}_{b}-as)^{3}-\frac{3m^{2}_{s}}{32\pi^{4}}\int^{1}_{a_{min}}da
\frac{(1-a)^{2}(6+a)}{a}(m^{2}_{b}-as)^{2}\nonumber\\&&-\frac{m_{s}\langle0|\bar{s}s|0\rangle}{2\pi^{2}}\int^{1}_{a_{min}}da
(1-a)(5-7a)(m^{2}_{b}-as)+\frac{m^{2}_{b}\langle0|g^{2}_{s}GG|0\rangle}{1152\pi^{4}}\int^{1}_{a_{min}}da
\frac{(1-a)^{3}(7+13a)}{a^{2}}\nonumber\\&&-\frac{\langle0|g^{2}_{s}GG|0\rangle}{384\pi^{4}}\int^{1}_{a_{min}}da(1-a)(2+a)(m^{2}_{b}-as)
+\frac{m^{2}_{s}\langle0|g^{2}_{s}GG|0\rangle}{384\pi^{4}}(1-a_{min})^{2}(6+a_{min})\nonumber\\&&+\frac{m_{s}\langle0|g_{s}\bar{s}\sigma\cdot Gs|0\rangle}{48\pi^{2}}\int^{1}_{a_{min}}da(1-4a+18a^{2})
-\frac{m_{s}\langle0|\bar{s}s|0\rangle\langle0|g^{2}_{s}GG|0\rangle}{144\pi^{2}M^{2}_{B}}(1-a_{min})(5-7a_{min})\nonumber\\&&
-\frac{m_{s}\langle0|\bar{s}s|0\rangle\langle0|g^{2}_{s}GG|0\rangle}{288\pi^{2}s}a_{min}(1+3a_{min}).
\end{eqnarray}

For the interpolating current $J^{\alpha_{1}\alpha_{2}}_{5/2,-,\Omega_{b},2,1,\lambda}(x)$,
\begin{eqnarray}
\Pi^{(OPE)\alpha_{1}\alpha_{2}\beta_{1}\beta_{2}}(p)=&&\not\!{p}g^{\alpha_{1}\alpha_{2}}g^{\beta_{1}\beta_{2}}(\int^{\infty}_{(m_{b}+2m_{s})^{2}}ds\frac{\rho(s)}{s-p^2}
+\frac{m_{s}\langle0|\bar{s}s|0\rangle\langle0|g^{2}_{s}GG|0\rangle}{1728\pi^{2}(m^{2}_{b}-p^{2})})
\nonumber\\&&+\mbox{other Lorentz structures},
\end{eqnarray}
where $\rho(s)$ is the QCD spectral density,
\begin{eqnarray}
\rho(s)=&&\frac{1}{288\pi^{4}}\int^{1}_{a_{min}}da
\frac{(1-a)^{3}(1+a)}{a}(m^{2}_{b}-as)^{3}-\frac{1}{288\pi^{4}}\int^{1}_{a_{min}}da\frac{(1-a)^{4}(1+2a)}{a}s(m^{2}_{b}-as)^{2}\nonumber\\&&
-\frac{1}{144\pi^{4}}\int^{1}_{a_{min}}da(1-a)^{5}s^{2}(m^{2}_{b}-as)-\frac{m^{2}_{s}}{32\pi^{4}}\int^{1}_{a_{min}}da
(1-a)^{2}(m^{2}_{b}-as)^{2}\nonumber\\&&+\frac{m^{2}_{s}}{48\pi^{4}}\int^{1}_{a_{min}}da
(1-a)^{3}s(m^{2}_{b}-as)-\frac{m_{s}\langle0|\bar{s}s|0\rangle}{18\pi^{2}}\int^{1}_{a_{min}}da
a(1-a)^{2}(3m^{2}_{b}-(a+1)s)\nonumber\\&&+\frac{m^{2}_{b}m_{s}\langle0|\bar{s}s|0\rangle}{18\pi^{2}}a_{min}(1-a_{min})^{3}
+\frac{m^{2}_{b}\langle0|g^{2}_{s}GG|0\rangle}{3456\pi^{4}}\int^{1}_{a_{min}}da
\frac{(1-a)^{3}(1+a)}{a}\nonumber\\&&-\frac{\langle0|g^{2}_{s}GG|0\rangle}{1152\pi^{4}}\int^{1}_{a_{min}}da(1-a)(2-a^{2})(m^{2}_{b}-as)
+\frac{\langle0|g^{2}_{s}GG|0\rangle}{6912\pi^{4}}\int^{1}_{a_{min}}da(1-a)^{2}(3-4a^{2})s\nonumber\\&&
+\frac{\langle0|g^{2}_{s}GG|0\rangle s}{3456\pi^{4}}(1-a_{min})^{4}+\frac{m^{2}_{b}\langle0|g^{2}_{s}GG|0\rangle}{3456\pi^{4}}a_{min}(1-a_{min})^{3}\nonumber\\&&
-\frac{\langle0|g^{2}_{s}GG|0\rangle s^{2}}{10368\pi^{4}M^{2}_{B}}(1-a_{min})^{5}-\frac{m^{2}_{s}\langle0|g^{2}_{s}GG|0\rangle}{3456\pi^{4}}(1-a_{min})^{2}(1-4a_{min})\nonumber\\&&
+\frac{m^{2}_{s}\langle0|g^{2}_{s}GG|0\rangle s}{3456\pi^{4}M^{2}_{B}}(1-a_{min})^{3}-\frac{m_{s}\langle0|g_{s}\bar{s}\sigma\cdot Gs|0\rangle}{72\pi^{2}}\int^{1}_{a_{min}}daa^{2}(1-2a)\nonumber\\&&-\frac{m_{s}\langle0|g_{s}\bar{s}\sigma\cdot Gs|0\rangle}{72\pi^{2}}\int^{1}_{a_{min}}daa(1-a)-\frac{m_{s}\langle0|g_{s}\bar{s}\sigma\cdot Gs|0\rangle}{432\pi^{2}}a_{min}(1-a_{min})(1-8a_{min})\nonumber\\&&-\frac{m^{2}_{b}m_{s}\langle0|g_{s}\bar{s}\sigma\cdot Gs|0\rangle}{108\pi^{2}M^{2}_{B}}a_{min}(1-a_{min})^{2}
-\frac{m_{s}\langle0|\bar{s}s|0\rangle\langle0|g^{2}_{s}GG|0\rangle}{1296\pi^{2}M^{2}_{B}}(1-a_{min})^{2}(4+a_{min})\nonumber\\&&
+\frac{m_{s}\langle0|\bar{s}s|0\rangle\langle0|g^{2}_{s}GG|0\rangle s}{1296\pi^{2}M^{4}_{B}}(1-a_{min})^{2}(5-4a_{min})-\frac{m_{s}\langle0|\bar{s}s|0\rangle\langle0|g^{2}_{s}GG|0\rangle s^{2}}{1296\pi^{2}M^{6}_{B}}(1-a_{min})^{3}\nonumber\\&&
-\frac{m_{s}\langle0|\bar{s}s|0\rangle\langle0|g^{2}_{s}GG|0\rangle}{5184\pi^{2}s}a_{min}
+\frac{m_{s}\langle0|\bar{s}s|0\rangle\langle0|g^{2}_{s}GG|0\rangle}{5184\pi^{2}M^{2}_{B}}a_{min}.
\end{eqnarray}

In the above equations, $a_{min}=m^{2}_{b}/s$ and $M^{2}_{B}$ is the Borel parameter.
\end{appendix}


\end{document}